\documentclass[modern]{aastex62}
\usepackage{epstopdf}
\usepackage{
  amsmath,       
  afterpage,     
  amssymb,       
  booktabs,      
  geometry,      
  graphicx,      
  multirow,      
  natbib,        
  physics
  }
 \usepackage{comment}
\usepackage{ifluatex}
\ifluatex
  \usepackage{pdftexcmds}
  \makeatletter
  \let\pdfstrcmp\pdf@strcmp
  \let\pdffilemoddate\pdf@filemoddate
  \makeatother
\fi
\usepackage{svg}  
\usepackage{float}
\usepackage{multirow}
\usepackage[caption=false, position=top]{subfig}
\usepackage{savesym}
\savesymbol{tablenum}
\usepackage[
  separate-uncertainty=true,
  multi-part-units=single, 
  bracket-numbers=false,
  angle-symbol-over-decimal=true]{siunitx}
\restoresymbol{SIX}{tablenum}
\usepackage{epstopdf}

\newcommand{\wes}{Department of Astronomy, Van Vleck Observatory, Wesleyan University, 96 Foss Hill Drive, Middletown, CT 06459, USA}

\begin{document}

\title{A Deep Search for Five Molecules in the 49 Ceti Debris Disk}
\shorttitle{49 Ceti Gas}

\author{Jessica Klusmeyer}
\affiliation{\wes}
\affiliation{NSF’s NOIRLab, 950 N. Cherry Ave., Tucson, AZ 85719, USA}

\author{A. Meredith Hughes}
\affiliation{\wes}

\author{Luca Matr\`{a}}
\affiliation{Center for Astrophysics $\vert$ Harvard \& Smithsonian, 60 Garden Street, Cambridge, MA 02138, USA}
\affiliation{Centre for Astronomy, School of Physics, National University of Ireland Galway, University Road, Galway, Ireland}

\author{Kevin Flaherty}
\affiliation{Department of Astronomy and Department of Physics, Williams College, Williamstown, MA 01267, USA}

\author{\'{A}gnes K\'{o}sp\'{a}l}
\affiliation{Konkoly Observatory, Research Centre for Astronomy and Earth Sciences, E\"otv\"os Lor\'and Research Network (ELKH), Konkoly-Thege Mikl\'os \'ut 15-17, H-1121 Budapest, Hungary}
\affiliation{Max Planck Institute for Astronomy, K\"onigstuhl 17, D-69117 Heidelberg, Germany}
\affiliation{ELTE E\"otv\"os Lor\'and University, Institute of Physics, P\'azm\'any P\'eter s\'et\'any 1/A, 1117 Budapest, Hungary}

\author{Attila Mo\'{o}r}
\affiliation{Konkoly Observatory, Research Centre for Astronomy and Earth Sciences, E\"otv\"os Lor\'and Research Network (ELKH), Konkoly-Thege Mikl\'os \'ut 15-17, H-1121 Budapest, Hungary}
\affiliation{ELTE E\"otv\"os Lor\'and University, Institute of Physics, P\'azm\'any P\'eter s\'et\'any 1/A, 1117 Budapest, Hungary}

\author{Aki Roberge}
\affiliation{Exoplanets \& Stellar Astrophysics Laboratory, NASA Goddard Space Flight Center, Code 667, Greenbelt, MD 20771, USA}

\author{Karin \"{O}berg}
\affiliation{Center for Astrophysics $\vert$ Harvard \& Smithsonian, 60 Garden Street, Cambridge, MA 02138, USA}

\author{Aaron Boley}
\affiliation{Department of Physics and Astronomy, University of British Columbia, Vancouver BC, Canada}

\author{Jacob White}
\affiliation{National Radio Astronomy Observatory, 520 Edgemont Rd.,
Charlottesville, VA, 22903, USA}
  \affiliation{Jansky Fellow of the National Radio Astronomy Observatory}

\author{David Wilner}
\affiliation{Center for Astrophysics $\vert$ Harvard \& Smithsonian, 60 Garden Street, Cambridge, MA 02138, USA}

\author{P\'{e}ter \'{A}brah\'{a}m}
\affiliation{Konkoly Observatory, Research Centre for Astronomy and Earth Sciences, E\"otv\"os Lor\'and Research Network (ELKH), Konkoly-Thege Mikl\'os \'ut 15-17, H-1121 Budapest, Hungary}
\affiliation{ELTE E\"otv\"os Lor\'and University, Institute of Physics, P\'azm\'any P\'eter s\'et\'any 1/A, 1117 Budapest, Hungary}

\begin{abstract}
Surprisingly strong CO emission has been observed from more than a dozen debris disks around nearby main-sequence stars.  The origin of this CO is unclear, in particular whether it is left over from the protoplanetary disk phase or is second-generation material released from collisions between icy bodies like  debris dust.  The primary unexplored avenue for distinguishing the origin of the material is understanding its molecular composition.  Here we present a deep search for five molecules (CN, HCN, HCO$^+$, SiO, and CH$_3$OH) in the debris disk around 49 Ceti.  We take advantage of the high sensitivity of the Atacama Large Millimeter/submillimeter Array (ALMA) at Band 7 to integrate for 3.2 hours at modest spatial (1'') and spectral (0.8\,km\,s$^{-1}$) resolution.  Our search yields stringent upper limits on the flux of all surveyed molecular lines, which imply abundances relative to CO that are orders of magnitude lower than those observed in protoplanetary disks and Solar System comets, and also those predicted in outgassing models of second-generation material.  However, if CI shielding is responsible for extending the lifetime of any CO produced in second-generation collisions, as proposed by \citet{2018kral}, then the line ratios do not reflect true {ice} phase chemical abundances, but rather imply that CO is shielded by its own photodissociation product, CI, but other molecules are rapidly photodissociated by the stellar and interstellar radiation field.  
\end{abstract}

%

\section{Introduction}
\label{section: introduction}

The dusty debris disks around main sequence stars are hallmarks of mature planetary systems.  The small quantity ($<<1$\,M$_\earth$) of dust in debris disks is generated by destructive processes involving grinding collisions of planetesimals \citep[][and references therein]{2018Hughes}.  Recent observations have revealed that the dust is commonly accompanied by detectable amounts of molecular gas \citep[e.g.,][]{1995zuc,moor2011,2017moor}.  The only { molecular} species so far detected is CO, which is detected most commonly around A- and B-type stars \citep{2016lieman} despite their harsh UV radiation fields that ought to photodissociate the gas on timescales much shorter than the age of the star \citep[e.g.,][]{2009visser,2013ApJkospal}.  Surprisingly, some of these systems exhibit CO emission comparable to that from the protoplanetary disks around pre-main sequence stars.  

Work to understand the origin and implications of the molecular gas in debris disks has centered around the question of whether the gas is effectively primordial, i.e., surviving from the protoplanetary disk phase and therefore H$_2$-dominated, or secondary, i.e., produced in destructive processes like the dust in debris disks and therefore H$_2$-poor.  The former category is sometimes referred to as ``hybrid" disks \citep[e.g.,][]{2013ApJkospal}, since it would imply the coexistence of primordial gas and secondary dust.  

Observationally, it is difficult to distinguish the origin of the gas on the basis of CO observations alone, although the weight of the evidence leans towards it being second-generation origin.  The low excitation temperatures are suggestive of a low H$_2$ gas mass, unlike the LTE conditions in protoplanetary disks \citep{2013ApJkospal}.  A secondary origin is also supported by the unresolved scale height in the 49 Ceti disk at a resolution of 0\farcs4 that suggests a high mean molecular weight \citep{Hughes2017}.  The presence of large amounts of CI gas in at least two systems has also been interpreted as evidence of second-generation origin \citep{2017higatomic}.  Certainly the asymmetries associated with individual systems like $\beta$ Pic and Fomalhaut are consistent with a second-generation origin \citep{dent2014,2017fomalhaut}.  On the other hand, the gas release rates required to sustain the large mass CO for the brightest systems, which can require vaporizing a small comet every few seconds if the CO is unshielded or a Hale-Bopp-size comet roughly once an hour \citep{2012Zuckerman,2013ApJkospal}, are difficult to reconcile with the putative mass reservoirs in debris disks and the high frequency with which gas is observed.  

Even if the origin of the gas is secondary, multiple mechanisms have been proposed to generate secondary gas.  {These mechanisms include} photodesorption \citep{gri07}, collisional vaporization of icy dust grains \citep{cze07}, and low-velocity collisions, either at the site of a resonant point with an unseen planet \citep{dent2014} or as the result of a major collision between planetary embryos \citep{jac14}, collisional desorption from comets \citep{2012Zuckerman}, or sublimation of comets in the inner regions of the system to explain absorption features \citep{beu90}.  Notably, \citet{kra17} developed a model based on the destruction of volatile-rich planetesimals that can explain the CO, C, and O content of most of the debris disks observed to date.  In that work they suggest that some of the higher-mass sources may be hybrid disks.  However, an extension of this basic model posits that even young debris disks with protoplanetary-level CO masses may in fact be composed of second-generation gas shielded by CI \citep{2018kral}.  

The chemical composition of the molecular gas is a relatively unexplored dimension of the properties of gas-bearing debris disks that can provide insight into the nature and origin of the gas.  Here we present the deepest search to date for five molecules other than CO in the debris disk around 49 Ceti.  49 Ceti is a 40 Myr \citep{2012Zuckerman} A-type star located 57.1\,pc from Earth \citep{2018gaia} that hosts one of the two brightest  (in CO) and best-studied of the gas-bearing debris disks.  The other system is $\beta$ Pictoris, which has an almost identical CO(3-2) line flux at a factor of $\sim$3 closer distance \citep{2017luca}.  Despite their similar CO fluxes, 49 Ceti is a better candidate for a molecular line survey because its larger gas densities are more readily capable of exciting emission from higher critical density molecules, and its higher disk-averaged excitation temperature (32\,K vs. 12\,K in $\beta$ Pic) is better for detecting lines with higher upper-level energies, thus implying that we will get the strongest constraints on the molecular abundances.  Its gas disk is also symmetric, unlike the $\beta$ Pic disk, and covers a smaller solid angle on the sky, making it easier to spatially average the emission to maximize sensitivity from faint lines.  Previous studies of the 49 Ceti system have spatially resolved the CO emission from the disk, revealing it to be axisymmetric and mostly consistent with Keplerian rotation about the central star between radii of $\sim$20 and $\sim$200\,au \citep{2008Hughes,Hughes2017}.  There are hints of an anomalous { (supersolar)} C/O ratio from {\it Herschel} spectroscopy \citep{rob14}, as well as evidence of active CO photodissociation with low H$_2$ abundances from the detection of strong CI emission \citep{hig17}.  

The five molecules selected for the search are CN, HCN, HCO$^+$, SiO, and CH$_3$OH.  The low cross-section of CN to photodissociation makes it more likely to survive in the harsh radiation field of an A star, and in fact models of second-generation gas based on comets with Solar System abundances indicate that CN should be about as bright as, if not brighter than, CO \citep{2018mat}.  It is also an excellent tracer of regions affected by UV radiation \citep{rod98}, and seems to be enhanced in evolved protoplanetary disks \citep{2014kast}.  CN is the daughter product of HCN photodissociation.  HCN tends to be bright in both protoplanetary disks and comets \citep{2002comet,2010oberkar}, although intriguingly it was not detected in previous, less-sensitive ALMA observations of the 49 Ceti system \citep{Hughes2017}.  HCO$^+$ is similarly bright in both disks and comets \citep{2016guillo,1997comet}, and tends to be produced in regions where water has been photodissociated \citep{mil04}.  It is also a sensitive tracer of ionization conditions in protoplanetary disks \citep{cle15}.  While SiO has never been detected at millimeter wavelengths in protoplanetary disks or comets, it would provide unequivocal evidence for grain-grain collisions as the source of the gas.  Methanol (CH$_3$OH) can be as abundant as CO in comets \citep{boc04,mum11}, but has only been detected once at a very low level in the nearest protoplanetary disk \citep{2016diskch3oh} and in an outbursting young star \citep{hoff18}.  

In Section~\ref{section: observations} we present the ALMA observations, and in Section~\ref{section: results} we describe the stringent upper limits we are able to place on the flux of all five molecules.  In Section~\ref{section: analysis} we apply a spectral shifting technique pioneered by \citet{2017luca} and similar to that of \citep{yen16} that takes advantage of the known CO velocity field to maximize any signal present in the noisy data.  We then discuss our results in the context of known and predicted molecular abundances for protoplanetary disks, comets, and second-generation gas in Section~\ref{section:discussion}, and summarize our conclusions in Section~\ref{section: conclusion}.  

%
\section{Observations}
\label{section: observations}

The observations (2017.1.00941.S, PI Hughes) utilized 48 of the 12-meter ALMA antennas in the most compact configuration available during Cycle 5. We observed using two spectral setups in Band 7, each executed on two separate dates.  On 2018 June 2 and June 6, the spectral setup targeted HCN(4-3) at a rest frequency of {354.5} GHz, CH$_3$OH 1(1,1)-0(0,0) at a rest frequency of 350.9 GHz and several hyperfine transitions of CN(3-2) with the strongest at a rest frequency of 340.2 GHz.  The other spectral setup observed on 2018 June 9 and 21 targeted SiO(8-7) at a rest frequency of 347.3\,GHz, CH$_3$OH 4(1,3)-3(0,3) at a rest frequency of 358.6 GHz and HCO$^{+}$(4-3) at a rest frequency of 356.7 GHz. The three spectral windows targeting line emission used a spectral resolution of 976\,kHz (0.8\,km\,s$^{-1}$) and bandwidth 1.875\,GHz to maximize { continuum} sensitivity. One spectral window in each setup was configured for continuum observations (at frequencies of 341.5 and 345.3\,GHz), with a bandwidth of 1.875 GHz and channel widths of 31.250 MHz { (27.5\,km\,s$^{-1}$)}.  The total on-source time was 1.6 hours per setup. 

Table~\ref{table:observational_params} gives the basic observational parameters for both spectral setups and all observing dates, including the  number of antennas, time on source, naturally weighted beam size ($"$), the coordinates of the phase center in RA and Dec, bandpass calibrators, flux calibrators, and gain calibrators.  Table~\ref{table:results} lists the rest frequency of each line (note that some were offset from the central frequency of the spectral window to satisfy constraints on the independently tunable intermediate frequencies (IFs).  

\begin{table}[ht]
    \centering
    \caption{49 Ceti Observational Parameters}
    \label{table:observational_params}
    \begin{tabular}{cccc}
        \hline
        \hline
       \multicolumn{2}{l} {$\#$ Antennas} &  \multicolumn{2}{c}{48} \\
      \multicolumn{2}{l}{ Baseline distance} & \multicolumn{2}{c}{12 - 250 m}\\
       \multicolumn{2}{l}{Time on Source} &  
       \multicolumn{2}{c}{3.2 h$^{a}$} \\
       \multicolumn{2}{l}{Naturally Weighted Beam Size} & \multicolumn{2}{c}{1\farcs1 x 0\farcs9} \\
       \multicolumn{2}{l}{ Phase center $\alpha$} &  \multicolumn{2}{c}{01:34:37.900573}\\  
        \multicolumn{2}{l}{Phase center $\delta$} & \multicolumn{2}{c}{-15:40:34.94661}\\  
        \bottomrule
        Dates & Bandpass Calibrator & Flux Calibrator & Gain Calibrator\\
        \bottomrule
        02-Jun-2018 & J0006-0623 & J0006-0623 &J0116-1136 \\
        06-Jun-2018 & J0006-0623 & J0006-0623 &J0116-1136 \\
        09-Jun-2018 & J0006-0623 & J0141-0928 &J0423-0120 \\
        21-Jun-2018 & J0423-0120 & J0423-0120 &J0141-0928 \\
        \bottomrule
    \end{tabular}
    \tablenotetext{a}{Divided into 1.6\,h per spectral setup}
\end{table}

The raw data were calibrated via the ALMA pipeline using the Common Astronomy Software Applications Package \citep[CASA;][]{2007mcmull} { version 5.5.0-149}.  We used the CASA task {\tt cvel} to convert the velocity axis from topocentric rest frame to the Kinematic Local Standard of Rest (LSRK).  We used {\tt UVCONTSUB} to fit a constant (a zeroth order polynomial) to channels more than 50 km s$^{-1}$ away from the known systemic velocity so that we could subtract the continuum.  We used {\tt TCLEAN} with natural weighting to produce channel maps for each molecule centered at the 2.78\,km\,s$^{-1}$ LSR systemic velocity of the 49 Ceti system \citep{Hughes2017} with pixel size 0\farcs15 covering an area 22\farcs5$\times$22\farcs5 on the sky.

%
\section{Results}
\label{section: results}

\begin{figure}[htb]
    \centering
    \includegraphics[width=1.0\textwidth,keepaspectratio]{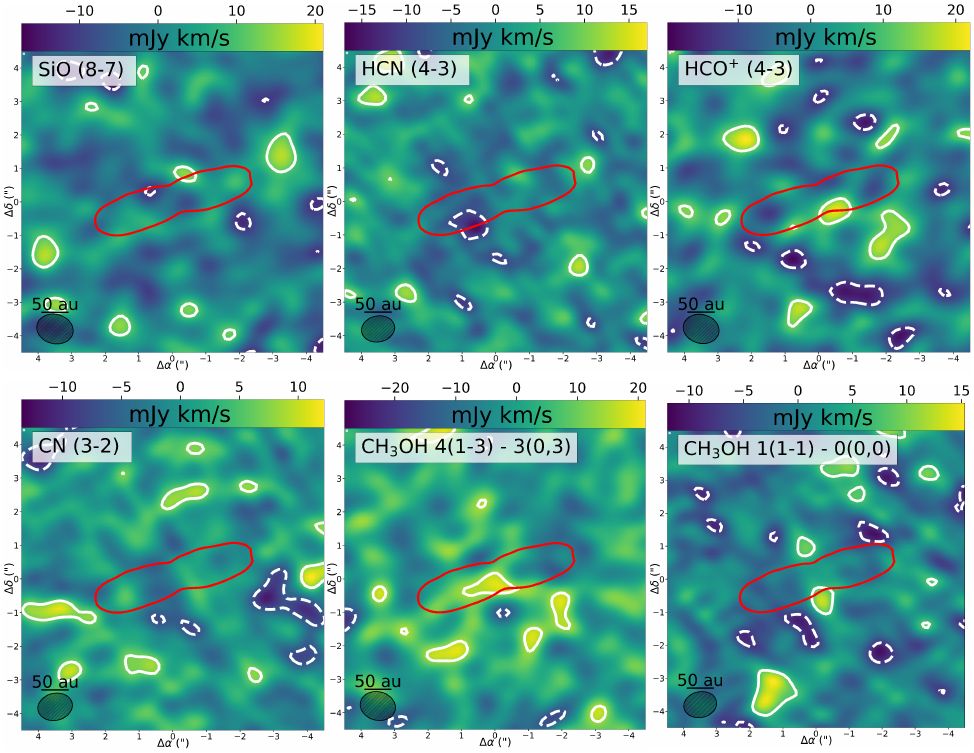}
 
 \caption{\label{fig:Moment0Maps} Moment 0 maps for all six molecular lines in the survey, integrated within $\pm5$\,km\,s$^{-1}$ of the 2.78\,km\,s$^{-1}$ systemic velocity of 49 Ceti.  The naturally weighted beam is indicated by the hatched ellipse located on the bottom left, below a 50\,au scale bar.  The red contour is an overlay of the CO(3-2) moment 0 map, indicating the spatial extent of the molecular line emission from the disk.  The white contours are $\pm2\sigma$, where $\sigma$ is the rms noise in the image: 4.6, 5.5, 6.4, 7.2, 3.7, and 3.4\,mJy\,km\,s$^{-1}$\,beam$^{-1}$, for HCN(4-3), SiO(8-7), HCO$^+$(4-3), CH$_{3}$OH 4(1,3)-3(0,3), CH$_3$OH 1(1,1)-0(0,0), and CN(3-2), respectively.  There is no statistically significant emission evident in any of the moment 0 maps.}
 
\end{figure}

We begin to search for emission from the five molecules by { using the} channel maps and zeroth-moment (velocity-integrated intensity) maps { generated} from the visibilities.  We use the CASA task {\tt immoments} to generate the moment 0 map, integrating channels within $\pm5$\,km\,s$^{-1}$ of the systemic velocity to ensure that all emission within the 8\,km\,s$^{-1}$ FWHM of the  CO line \citep{2008Hughes} is incorporated.  The moment 0 maps are displayed in Figure~\ref{fig:Moment0Maps} and show no significant gas emission.

CH$_{3}$OH and CN have multiple transitions in their spectral windows.  CH$_{3}$OH has two transitions, 1(1,1)-0(0,0) and 4(1,3)-3(0,3), in two different spectral windows.  CN exhibits many different hyperfine transitions within the same spectral window, all of which are listed in Table \ref{table:CN}.  For our initial reconnaissance, we shifted all of the hyperfine lines to a common velocity channel and added them together in both the { disk-integrated, shifted} spectra produced from the channel maps { (see Section~\ref{section: analysis} for details)}, and in the moment 0 map, but did not detect any emission.  For all subsequent analyses, we generate channel maps, spectra, and moment 0 maps for each methanol transition separately, and only for the strongest (highest intensity) of the CN transitions.

\begin{table}[ht!]
    \centering
    \caption{CN Hyperfine Transitions}
    \label{table:CN}
    \begin{tabular}{*3c}
        \toprule
         Rest Frequency$^a$ (GHz)& Transition & CDMS/Lovas Intensity$^a$ \\  
        \midrule

339.44677 	& N= 3-2, J=5/2-5/2, F=3/2-3/2	& -3.674 \\
339.45999   & N= 3-2, J=5/2-5/2, F=3/2-5/2	& -4.3924 \\
339.46263 	& N= 3-2, J=5/2-5/2, F=5/2-3/2	& -4.3831 \\
339.47590 	& N= 3-2, J=5/2-5/2, F=5/2-5/2	& -3.5256 \\
339.49321 	& N= 3-2, J=5/2-5/2, F=5/2-7/2	& -4.3777 \\
339.49928 	& N= 3-2, J=5/2-5/2, F=7/2-5/2	& -4.3598 \\
339.51663 	& N= 3-2, J=5/2-5/2, F=7/2-7/2	& -3.3239 \\
339.99225 	& N= 3-2, J=5/2-3/2, F=3/2-5/2	& -4.4392 \\
340.00809 	& N= 3-2, J=5/2-3/2, F=5/2-5/2	& -3.0613 \\
340.01960 	& N= 3-2, J=5/2-3/2, F=3/2-3/2	& -3.0625 \\
340.03154 	& N= 3-2, J=5/2-3/2, F=7/2-5/2	& -2.1437 \\
340.03540 	& N= 3-2, J=5/2-3/2, F=3/2-1/2	& -2.5692 \\
340.03540 	& N= 3-2, J=5/2-3/2, F=5/2-3/2	& -2.3442 \\
340.03540	& N= 3-2, J=5/2-3/2, F=3/2-1/2	& -2.5692 \\
340.03540 	& N= 3-2, J=5/2-3/2, F=5/2-3/2	& -2.3442 \\
340.24777 	& N= 3-2, J=7/2-5/2, F=7/2-5/2	& -2.1495 \\
340.24777$^{*}$& N= 3-2, J=7/2-5/2, F=9/2-7/2	&  -2.0159 \\
340.24854 	& N= 3-2, J=7/2-5/2, F=5/2-3/2	& -2.2887 \\
340.26177 	& N= 3-2, J=7/2-5/2, F=5/2-5/2	& -3.2027 \\
340.26494 	& N= 3-2, J=7/2-5/2, F=7/2-7/2	& -3.2039 \\
340.27912 	& N= 3-2, J=7/2-5/2, F=5/2-7/2	& -4.8867 \\          
 \bottomrule
 $^{*}$ strongest transition     \\
    \end{tabular}
    \tablenotetext{a}{https://www.cv.nrao.edu/php/splat/}
\end{table}

%
\section{Analysis}
\label{section: analysis}

To obtain the most sensitive upper limits on the flux density of the molecules, we utilize a spatial-spectral shifting technique pioneered by \citet{2017fomalhaut}.  The original technique uses a model velocity field to identify the expected location of flux within the 3-D position-position-velocity cube and shift the expected peak of the spectrum of each spatial pixel to a common velocity, thereby concentrating signal in a few channels.  This technique substantially increases the signal-to-noise ratio (by concentrating signal rather than changing the noise level), as evidenced by the detection of a small quantity of CO in the disk around Fomalhaut that was not evident in the unprocessed channel or moment 0 maps \citep{2017fomalhaut}.  

\begin{figure}[ht!]
\centering
\includegraphics[width=0.6\textwidth,height=0.5\textheight,keepaspectratio]{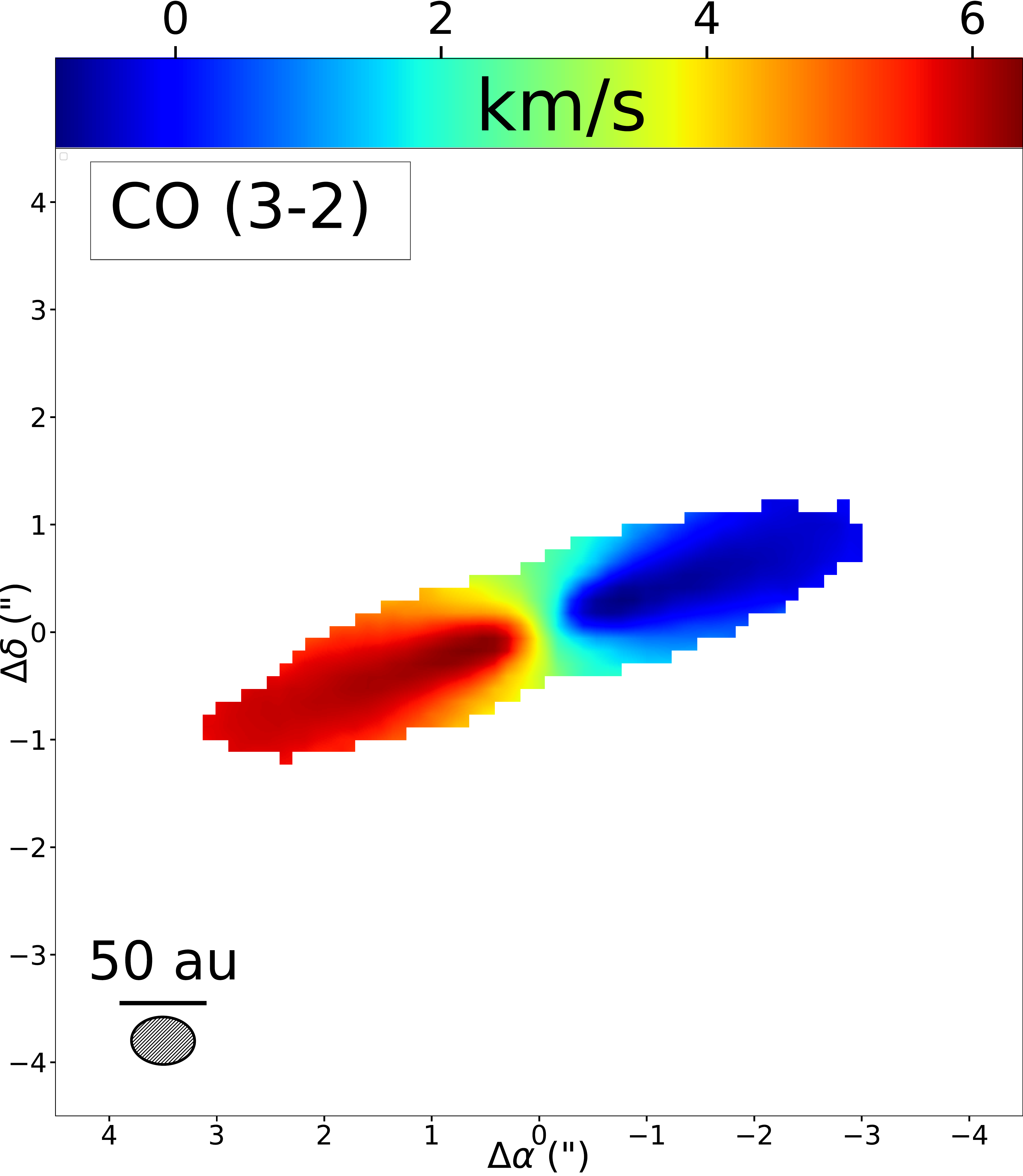}
\caption{A moment 1 map of CO(3-2) in the 49 Ceti debris disk generated from data presented in \citet{Hughes2017}. We use this observed intensity-weighted velocity field as the input for the spectral shifting technique drawn from \citet{2017fomalhaut}.}
\label{Moment1co}
\end{figure}

We modify the technique slightly to utilize an observed velocity field rather than a model of a flat Keplerian disk.  Previous observations of the 49 Ceti disk have demonstrated that 80\% of the CO(3-2) flux is well described by an axisymmetric power-law disk in Keplerian rotation \citep{Hughes2017}, although the remaining 20\% of the flux shows intriguing deviations from axisymmetry.  We therefore use the moment 1 map of the CO(3-2) emission as the ``model" velocity field, and shift each pixel according to the intensity-weighted velocity of CO at that location.  Applying the method to the CO(3-2) data cube boosts the signal-to-noise ratio of the line by a factor of 2-3 over a flat Keplerian model for 49 Ceti \citep{2018lambros}, although { extending that result to the current data} relies on the assumption that the targeted molecules share a common spatial distribution and velocity field with CO.  Figure~\ref{Moment1co} shows the CO(3-2) moment 1 map generated from the data in \citet{Hughes2017}, using the CASA task {\tt immoments} and excluding pixels below three times the rms noise.

\begin{figure*}[ht!]
\begin{center}
\includegraphics[width=1.0\textwidth,keepaspectratio]{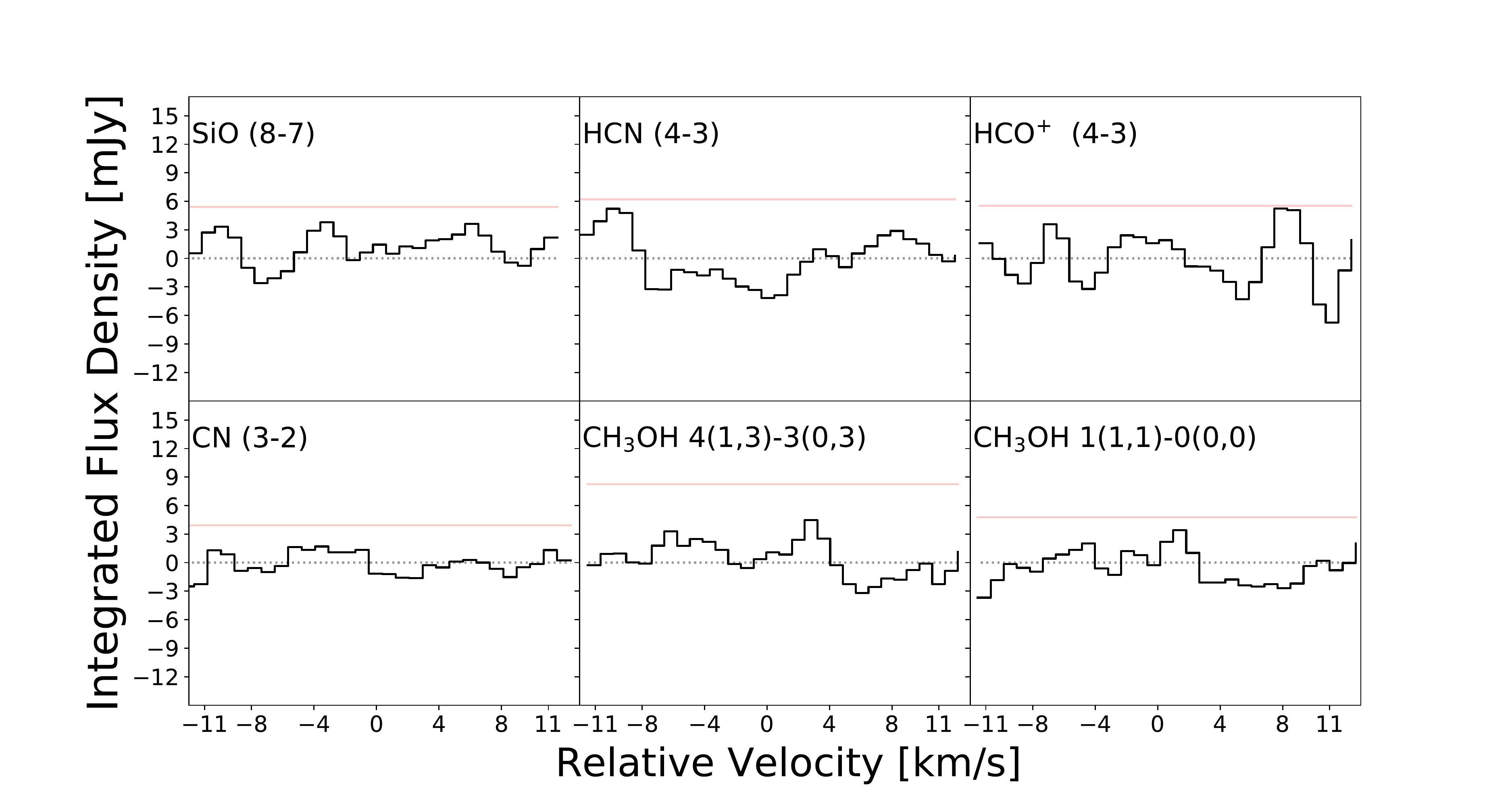}

\end{center}

\caption{\label{fig:shiftspec}The shifted spectra for all five targeted molecules in the 49 Ceti system.  { The x-axis shows the velocity relative to the LSR velocity of 49 Ceti (so that 0\,km\,s$^{-1}$ is the expected velocity of the emission).}  Each spectrum was created utilizing the shifted channel maps, which were then integrated over the area of the disk enclosed by the 3$\sigma$ contours of the CO(3-2) moment 0 map.   The CN spectrum shows only the strongest (highest intensity) of the hyperfine transitions.  The red horizontal line denotes the 3$\sigma$ flux level for each spectrum (three times the rms noise for the spectrum, calculated across at least 100 channels around the line within each spectral window).  There is no significant ($>3\sigma$) emission near a velocity of 0\,km\,s$^{-1}$ for any of the observed lines.}
\end{figure*}

We applied this technique to our five molecules, subtracting the observed CO(3-2) gas velocity from the spectrum of each molecule in each pixel to shift any flux to a common zero-velocity channel.  We then use the MIRIAD\footnote{Multichannel Image Reconstruction, Image Analysis and Display; \citep{sault95}} task {\tt cgcurs} to integrate the flux within the mask defined by the 3$\sigma$ contours of the CO(3-2) moment 0 map.  For molecules with multiple transitions (i.e., the CN hyperfine transitions listed in Table~\ref{table:CN} and the two methanol transitions), we initially combined all the transitions by independently shifting and then subsequently adding together the 3-D position-position velocity cube for each transition before integrating the spectrum; however since no emission was detected we ultimately reported the upper limit for only the strongest transition in each spectral window.  The resulting shifted spectra for all molecules are displayed in Figure~\ref{fig:shiftspec}.   None of the spectra exhibit statistically significant { ($>3\sigma$)} emission { near the expected velocity}.  

{ Table~\ref{table:results} lists the 3$\sigma$ upper limit on the integrated flux for each line, calculated within the area enclosed by the 3$\sigma$ contours of the shifted CO moment { 0} map and integrated over $\pm$2.0\,km\,s$^{-1}$ along the velocity axis. We restrict the velocity to this range because \citet{2018lambros} demonstrated that applying the spectral shifting technique to the CO data from 49 Ceti narrowed the line considerably from its original 8\,km\,s$^{-1}$ width, but that a range of $\pm2$\,km\,s$^{-1}$ was still required to detect all of the shifted flux.  The upper limit is calculated by measuring the rms noise in the moment 0 map, multiplying by the square root of the number of synthesized beams enclosed within the 3$\sigma$ contours of the shifted CO moment map (the area within which the integrated flux is measured), and then multiplying by three to obtain the 3$\sigma$ upper limit.}

\begin{table}[H]
    \centering
    \caption{Upper limits (3$\sigma$) on molecular gas emission from the 49 Ceti disk}
    \label{table:results}
    \begin{tabular}{*4c}
        \toprule
         Species& Rest Frequency$^a$  & Upper Limit & Gas Mass  \\  
        & GHz &\  mJy\,km\,s$^{-1}$ & $M_\oplus$\\
        \midrule
        CN(3-2)&340.248*& {25}&  {$<$2.0 $\times$ 10$^{-8}$ }\\
        SiO(8-7)&347.3305 & {34}&{$<$8.3 $\times$ 10$^{-9}$} \\
        HCN(4-3)&354.5054 & {31}  &{$<$1.7 $\times$ 10$^{-9}$}       \\
        HCO$^{+}$(4-3)&356.7342 & {37} &{$<$1.3 $\times$ 10$^{-9}$  }   \\

        \textnormal{C}H$_{3}$OH **&350.9051 & {22} &{$<$1.5 $\times$ 10$^{-7}$} \\
        \textnormal{C}H$_{3}$OH ***&358.6058 & {56} & {$<$7.6 $\times$ 10$^{-7}$} \\
        \bottomrule
        & *N=3-2, J=7/2-5/2, F=9/2-7/2 **1(1,1)-0(0,0) &  ***4(1,3)-3(0,3)  \\
        \bottomrule       
    \end{tabular}
    \tablenotetext{a}{https://www.cv.nrao.edu/php/splat/}
\end{table}

Even with the signal boost from the spatial-spectral shifting technique, we do not detect any gas emission from the five molecules.  We convert the flux upper limits (calculated based on the noise level {in the shifted moment 0 maps with the velocity range restricted to $\pm2$\,km\,s$^{-1}$}) to mass upper limits, assuming optically thin emission and local thermodynamic equilibrium (LTE) and adopting the disk-averaged excitation temperature of T$_\mathrm{ex}$ = 32\,K from the CO line ratio in \citet{Hughes2017}, using the following relationships:

\begin{align}
M = \frac{4\pi}{h\nu_{0}}\frac{F_{J_{u} - J_{l}} \textnormal{m}_{mol} \textnormal{d}^{2}}{A_{J_{u}- J_{l}
}X_{u}}
\end{align}
\begin{align}
X_{u} = \frac{g_u}{Q(T)}e^{-E_u/k T_{ex}}
\end{align}

where $m_\mathrm{mol}$ is the molecular mass, $d$ is the distance to the source, $\nu_{0}$  is the rest frequency for each transition taken from the Splatalogue database, $A_{J_{u}-J_{l}}$ is the Einstein coefficient { (also taken from Splatalogue)}, X$_{u}$ represents the fraction of molecules in the upper energy state, and $F_{J_{u}-J_{l}}$ is the upper limit on the flux density for a given transition from Table~\ref{table:results}. The mass of each molecule, $m_\mathrm{mol}$, the degeneracy of the upper state $g_u$, the energy of the upper state $E_u$, and the tabulated partition function at a temperature of 37.5K (the closest to our assumed excitation temperature) $Q(T)$, were taken from the Cologne Database for Molecular Spectroscopy (CDMS).  
Table~\ref{table:molecular_params} summarizes the parameters for each molecule, and Table~\ref{table:results} lists the corresponding upper limit on the gas mass of each molecule.

\begin{table}[ht!]
\center
    \caption{Molecular Parameters}
    \label{table:molecular_params}
    \scalebox{0.8}{%
    \begin{tabular}{*6c}
        \toprule
    
        Species & Mass &    A (s$^{-1}$)$^{a}$ & $E_u$ (cm$^{-1}$)$^b$ & g$_{u}^b$ & Q(T)$^b$ \\ 
         &(Da) & (mJy km s$^{-1}$) & & & \\
        \midrule
          CN             &  26.00362$^{b}$&         4.1313 $\times$ 10$^{-4}$& 22.7028 & 10 & 84.7308 \\
         SiO            &  43.97184$^{b}$&  2.2030 $\times$ 10$^{-3}$& 52.1397 & 17 & 36.3268\\

        HCN            &  27.01090$^{a}$ &    2.0605 $\times$ 10$^{-3}$& 29.5633 & 27 & 53.9106\\
                  
         HCO$^{+}$      &   29.00274$^{c}$&   3.5722 $\times$ 10$^{-3}$ &   29.7491 & 9 & {17.8601} \\

         CH$_{3}$OH     &   32.02622$^{b}$&        3.3148 $\times$ 10$^{-4}$ & 11.7049 & 12 & 920.963739\\
         CH$_{3}$OH     &   32.02622$^{b}$&     1.3193 $\times$ 10$^{-4}$&   30.7644 & 36 & 920.963739\\

        \bottomrule

    \end{tabular}}
    \tablenotetext{b}{www.chemspider.com}
    \tablenotetext{c}{~~~~cdms.astro.uni-koeln.de \citep{2016JMoSpCDMS}}
    \tablenotetext{d}{\citet{kim2019}}
\end{table}\vfill


%

\section{Discussion}
\label{section:discussion}

To date, CO remains the only firmly detected gas-phase molecule in debris disks \citep{2018Hughes}.  The upper limit on HCN(4-3) from our line survey is deeper than a previous search for HCN in the 49 Ceti disk \citep{Hughes2017}, by a factor of about 2.5 in rms noise, which is enhanced to { $\sim5\times$} by our use of the spectral shifting technique.   \citet{2018mat} conducted a Submillimeter Array (SMA) and ALMA survey for CN, HCN, HCO$^+$, N$_2$H$^+$, H$_2$CO, H$_2$S, CH$_3$OH, SiO, and DCN in the $\beta$ Pic disk that also returned upper limits.  When applying a simple scaling relationship for the differences in distance, frequency, { and Einstein coefficient} to the flux upper limits, our survey is roughly { 7--13}$\times$ more sensitive to CN, HCN, HCO$^+$, { and methanol}.  The emerging picture seems to be that the abundance of molecules other than CO is surprisingly low in debris disks, even for those molecules that exhibit relatively high abundances in (1) protoplanetary disks, (2) solar system comets, and (3) models of secondary gas emission.  Here we explore how the gas composition implied by our upper limits compares with the abundance ratios of those three categories of objects.  

\subsection{Comparison with Observed Molecular Abundances of Protoplanetary Disks and Comets}

Protoplanetary disks exhibit a rich chemistry that has been studied both observationally and theoretically, with the field undergoing a particularly rapid expansion recently as ALMA has reached maturity \citep{hen13,dut14,ber18}.  All of the molecules in our survey, with the exception of SiO, have been detected in one or more protoplanetary disks.  

Models of protoplanetary disk chemistry predict a rich abundance of molecular species, as a result of stellar and interstellar radiation, cosmic rays, and grain surface chemistry \citep[e.g.,][]{1999Aikawa,1999yuri}.  HCO$^+$ and HCN are particularly abundant, both in the models and subsequent observations.  In a survey of eight T Tauri disks and four Herbig Ae disks with the SMA, HCO$^+$(3-2) emission was detected towards all of the surveyed objects and HCN(3-2) was at least tentatively detected towards 10 of the 12 sources \citep{2010oberkar,2011chemoberg}.  Typical rms noise levels in the survey were of order $\sim0.1$\,Jy\,km\,s$^{-1}$ at distances of $\sim$140\,pc { (although V4046 Sgr was closer)}.  If we scale those thresholds by $\nu^2$ (assuming optically thick emission) and by $d^2$ to account for the difference in distance to the 49 Ceti system, then our upper limits on the integrated HCO$^+$(4-3) and HCN(4-3) flux from 49 Ceti are about a factor of 50 below the sensitivity of the \citet{2010oberkar,2011chemoberg} survey.  The CO(2-1) emission from 49 Ceti \citep[$3.87\pm0.41$\,Jy\,km\,s$^{-1}$,][]{moor2019} is 26 times lower than the average for the disks in the \citet{2011chemoberg} survey, and 10 times lower than the faintest disk in the survey.  { Looking at it another way, the ratio of { HCO$^+$/CO} flux within a given band varies in the \citet{2010oberkar,2011chemoberg} sample between { 0.11 and 0.59}, while the ratio for 49 Ceti at Band 7 is { $<0.006$}.  HCN is less conclusive, since there are two upper limits within the sample of 12, but for sources in which it is detected the HCN/CO ratio ranges from 2.0 to 18, while the ratio for 49 Ceti is $<0.005$.} A protoplanetary composition for the 49 Ceti disk is not supported by the low limits on both HCO$^+$ and HCN compared to CO emission. 

The nondetections of CH$_3$OH and CN are less relevant for comparison with a protoplanetary disk composition.  While \citet{2014protopredic} predicted detectable CH$_3$OH emission from protoplanetary disks based on models of grain-surface chemistry, nonthermal desorption, and irradiation of grain mantles, subsequent observations yielded lower abundances than expected by two orders of magnitude \citep{2016diskch3oh}.  Observations of the nearest and brightest Herbig Ae disk (around HD 163296) have demonstrated an even lower CH$_3$OH/H$_2$CO ratio than in the TW Hya system (the closest protoplanetary disk and the only non-outbursting source in which methanol has been detected to date), suggesting that abundances may be suppressed even further in the radiation field of an A star.  We therefore would not expect to detect CH$_3$OH in the 49 Ceti system { even} if its abundance relative to CO were protoplanetary. Similarly, a recent ALMA survey of molecular lines from protoplanetary disks in Lupus returned detections of CN(3-2) in only 38\% of surveyed sources \citep{2019cnlupus}.  Typical upper limits were of order $\sim100$\,mJy\,km\,s$^{-1}$, however, which is about a factor of { 30} above our upper limits when scaled to the appropriate distance.  This result is similar to previous surveys showing that $\sim$50\% of bright protoplanetary disks exhibit detectable CN at modest sensitivity \citep[e.g.,][]{gui13}.  Our nondetection of CN is therefore inconclusive as a probe of protoplanetary composition. 

Solar system comets serve as a foundation for predicting the likely composition of second-generation gas in extrasolar systems, although they are imperfect comparison objects for many reasons -- foremost that most Solar System comets are observed while sublimating at distances of a few au from the Sun, whereas the gas in extrasolar debris disks originates from tens to hundreds of au from the central star.  The radiation field of an A star differs markedly from that of the Sun, and the mechanism behind the gas release may differ as well (collisional vs. radiative).  

In a millimeter-wavelength survey of the composition of 24 comets with the Institut de Radioastronomie Millimetrique 30m, James Clerk Maxwell Telecope, Caltech Submillimeter Observatory, and Swedish-ESO Submillimeter Telescope, HCN was detected in all 24 comets with an abundance relative to CO of order $\sim$0.1-1 \citep{2002comet}.  HCN lines in this survey were typically as bright or brighter than the CO lines observed in the same band.  These compositional relationships are also borne out across larger and more modern samples of comets \citep{mum11,boc17}.  In the 49 Ceti system, the HCN(4-3) upper limit of {31}\,mJy\,km\,s$^{-1}$ from this work implies an HCN(4-3)/CO(3-2) line ratio of {$< 0.0052$}, using the CO(3-2) flux of $6.0\pm 0.1$\,Jy\,km\,s$^{-1}$ from \citet{Hughes2017}.  This limit is well below the observed range of HCN(4-3)/CO(3-2) line ratios for Solar System comets, which span values of { $\sim$1-20, although for most systems CO was not detected at all with lower limits of order $\gtrsim 10$} \citep[][and references therein]{2002comet}.  HCO$^+$ has also been detected in comets and may be particularly prevalent in regions where water has been photodissociated \citep{mil04}.  If the 49 Ceti disk exhibited the same abundance patters as Solar System comets, we would therefore strongly expect to detect HCN, and would likely detect methanol and HCO$^+$.  The lack of detection of any of these molecules suggests that we need to dig deeper into modeling of second-generation gas to understand why the abundance ratios indicated by our observations differ so markedly from {\it both} { protoplanetary} disks and comets. 

\subsection{Models of Second-Generation Gas Production}

In this section we explore models of second-generation gas production, beginning with an estimate of the CO photodissociation and survival timescale assuming only shielding by H$_2$ and self-shielding by CO (Section~\ref{COphd}).  We then extend the analysis to CN, which is the molecule in our survey that has the best combination of a long survival timescale against photodissociation and intrinsically strong molecular transitions (Section~\ref{CNphd}).  Finally, we derive constraints on the HCN/(CO+CO$_2$) outgassing rate ratios (Section~\ref{HCN_rate}).  The latter two sections follow the methodology derived in \citet{2018mat}, applying it to these new and more stringent constraints for the specific case of the 49 Ceti system.

\subsubsection{CO photodissociation and survival timescale}
\label{COphd}
As ALMA has matured and detections of gas in debris disks have started piling up \citep[e.g.,][]{1995zuc,moor2011,moo15,2017moor,dent2014,gre16,2016lieman,2016diskwcomets,mar17,2017fomalhaut}, there have been a number of advances in models of second-generation gas production in debris disks.  There are several proposed mechanisms, including collisional gas release \citep{2012Zuckerman,dent2014}, thermal desorption \citep{cze07}, and photodesorption \citep{gri07}.  Recent modeling efforts have primarily focused on collisional gas release from icy planetesimals as the origin of the outgassing responsible for the observed molecular lines \citep{2017luca,kra17}.  In this section we apply the model presented in \citet{2017luca} to our observations of the 49 Ceti disk to investigate the origin of the gas.  

We first estimate the photodissociation timescale for CO using models of the temperature and density structure of the CO disk presented in \citet{Hughes2017}.  We calculate { half} the vertical column density through the disk.  We perform this calculation for both scenarios investigated in the paper: a primordial scenario in which the CO/H$_2$ abundance is assumed to equal 10$^{-4}$, and a second-generation scenario in which the CO/H$_2$ abundance is assumed to equal 1.  We also scale up the mass of the \citet{Hughes2017} best-fit models to match the total CO mass of (1.11 $\pm 0.13) \times 10^{-2}$ M$_{\oplus}$ derived by \citet{moor2019}.  In the primordial scenario  (where the gas is in LTE) the CO column density is  { 2.0} $\times$ 10$^{16}$ cm$^{-2}$ and the H$_2$ column density is { 2.0} $\times$ 10$^{20}$ cm$^{-2}$, whereas in the second-generation scenario (where the conditions are non-LTE), both CO and H$_{2}$ have column densities of { 3.6} $\times$ 10$^{16}$ cm$^{-2}$. 

Since the dust in the 49 Ceti system is optically thin, we assume negligible dust attenuation. We then estimate the photodissociation timescale using the following relationships, following \citet{fla19}: 
\begin{align}
k&=\chi k_{0} \Theta e^{-\gamma A_{v}} \\
t&=1/k
\end{align}
In these equations, $\chi$ is the radiation field in units of the Draine field \citep[one Draine field is $\sim$1.7 Habings, where a Habing is $1.6\times10^{-3}$\,erg\,cm$^{-2}$\,s$^{-1}$;][]{dra78}. We consider only the interstellar radiation field (ISRF), and ignore contributions from the stellar radiation field (which would only decrease the photodissociation timescale).  $k_{0}$ is the unattenuated photodissociation rate in units of photodissociations per second for CO molecules.  We used $k_{0}$ = 2.590 $\times$ 10$^{-10}$\,s$^{-1}$ from Table 6 of \cite{2009visser}, since Table 6 assumes an excitation temperature of 50\,K which is closest to the disk-averaged excitation temperature of 32\,K derived by \citet{Hughes2017}. $\Theta$ is the attenuation factor for self-shielding and H$_{2}$ shielding, tabulated in \citep{2009visser} for different values of CO and H$_2$ column density.  $\gamma$ is the UV extinction relative to visual extinction ($\approx$2 for small dust grains).  The photodissociation timescale is then $1/k$.  We consider the simple case with no dust attenuation throughout the disk ($A_V = 0$), using the value of $\Theta$ corresponding to the column density, which amounts to an assumption that the photodissociation timescale is set by the most heavily shielded molecules in the disk \citep{2009visser}.  We calculate a photodissociation timescale of  { $\sim$6,000\,yr} for the primordial scenario and { $\sim$3,300\,yr} for the second-generation case (due to the differences in column density, which correspond to different values of $\Theta$).  These timescales should be considered upper limits, since we have neglected stellar radiation, which could contribute up to the same amount of radiation as the ISRF depending on the optical thickness of the radial gas column.  

These results present a familiar dilemma: the photodissociation timescale is so short compared to the age of the star that a primordial origin for the gas is unlikely, but the combination of large CO mass and short photodissociation timescale requires an uncomfortably high gas release rate from large bodies to sustain the observed gas \citep[see, e.g.,][]{2013ApJkospal}.  Recent work by \citet{2018kral} suggests that the solution to the dilemma may lie in shielding of CO by neutral carbon in the disk, which is not included in the standard calculation of the photodissociation timescale that we performed, but which can extend the lifetime of CO by orders of magnitude.  Observations by \citet{2017higatomic} demonstrated substantial reservoirs of neutral carbon in the disk, as expected due to rapid initial photodissociation of CO, and a detailed analysis by \citet{moor2019} demonstrated that the amount of CO is consistent with the \citet{2018kral} model given the observed quantity of CI in the 49 Ceti system.  These results provide evidence for a secondary origin for CO in the disk.  They are supported by the unresolved vertical height of the CO disk, which is smaller than expected for a primordial composition and implies a mean molecular weight more consistent with CO (rather than H$_2$) as the dominant constituent \citep{Hughes2017}.  However, it is worth noting that the long photodissociation timescale of CO implied by CI shielding would apply equally to CO of primordial or secondary origin. 

\subsubsection{CN photodissociation, mass, and survival timescale}
\label{CNphd}
{ While the LTE estimates provide a ballpark value for the conversions between flux and mass, if the H$_2$ density is low, then it is likely that at least some species are not in LTE.}  Moving on to molecules other than CO, we can convert the upper limits in our survey to gas masses via the non-LTE calculation described in \citet{2018mat}. We focus on CN, since for Solar System comet-like compositions it is expected to be the most favourable for mm-wave detection \citep{2018mat}, compared to other molecules expected in a cometary environment (like the observed CH$_3$OH and HCN). This is largely due to CN having the best combination of a long survival timescale against photodissociation, and intrinsically strong molecular transitions. We do not consider SiO gas as we are not aware of its detection in Solar System comets, and HCO$^+$ because connecting its gas mass to parent species in exocometary ice would depend on a complex network involving chemistry and ionization balance, which is beyond the scope of this paper.

\begin{figure}[ht!]
\centering
\includegraphics[width=0.7\textwidth,keepaspectratio]{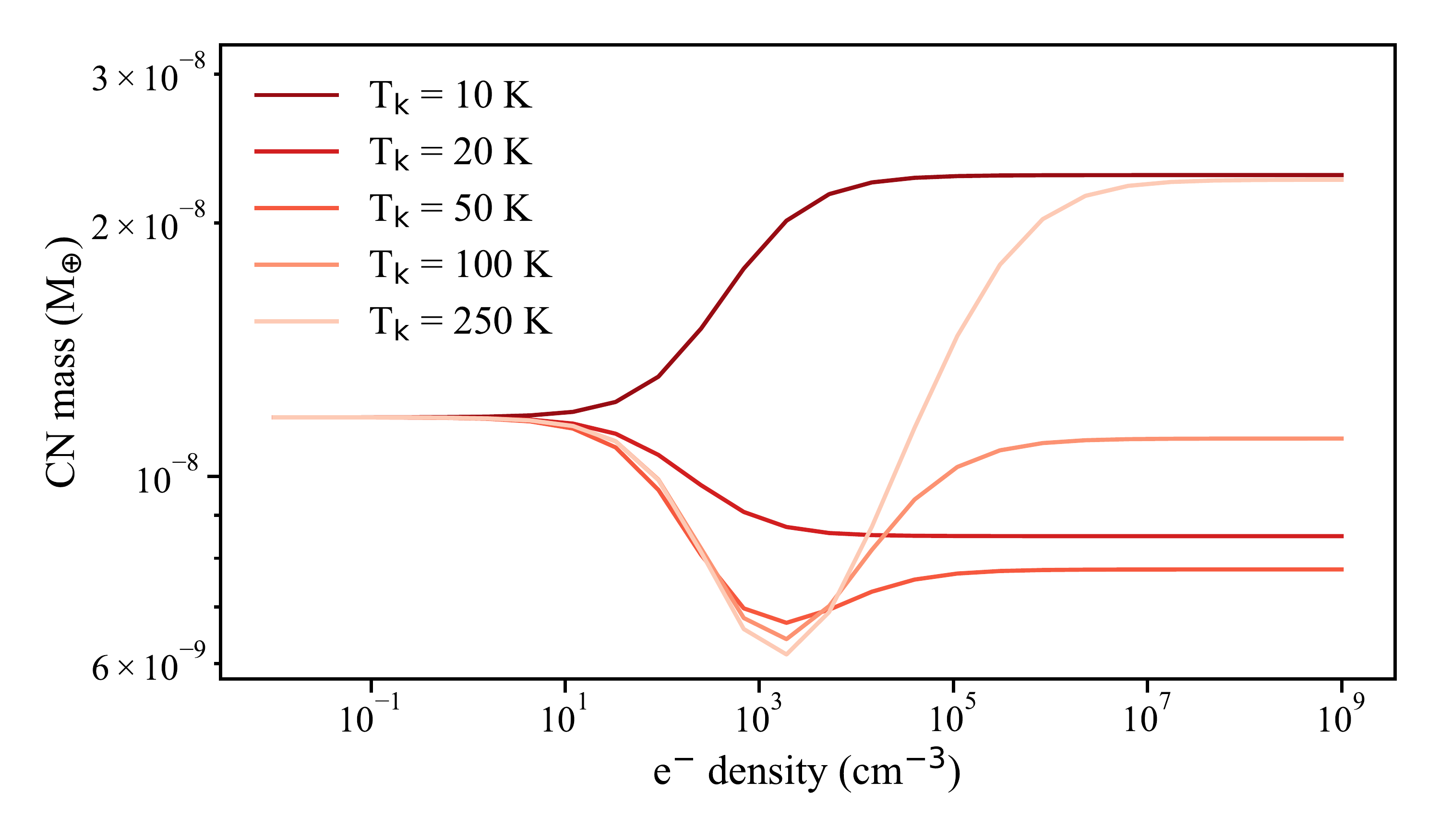}
\includegraphics[width=0.7\textwidth,keepaspectratio]{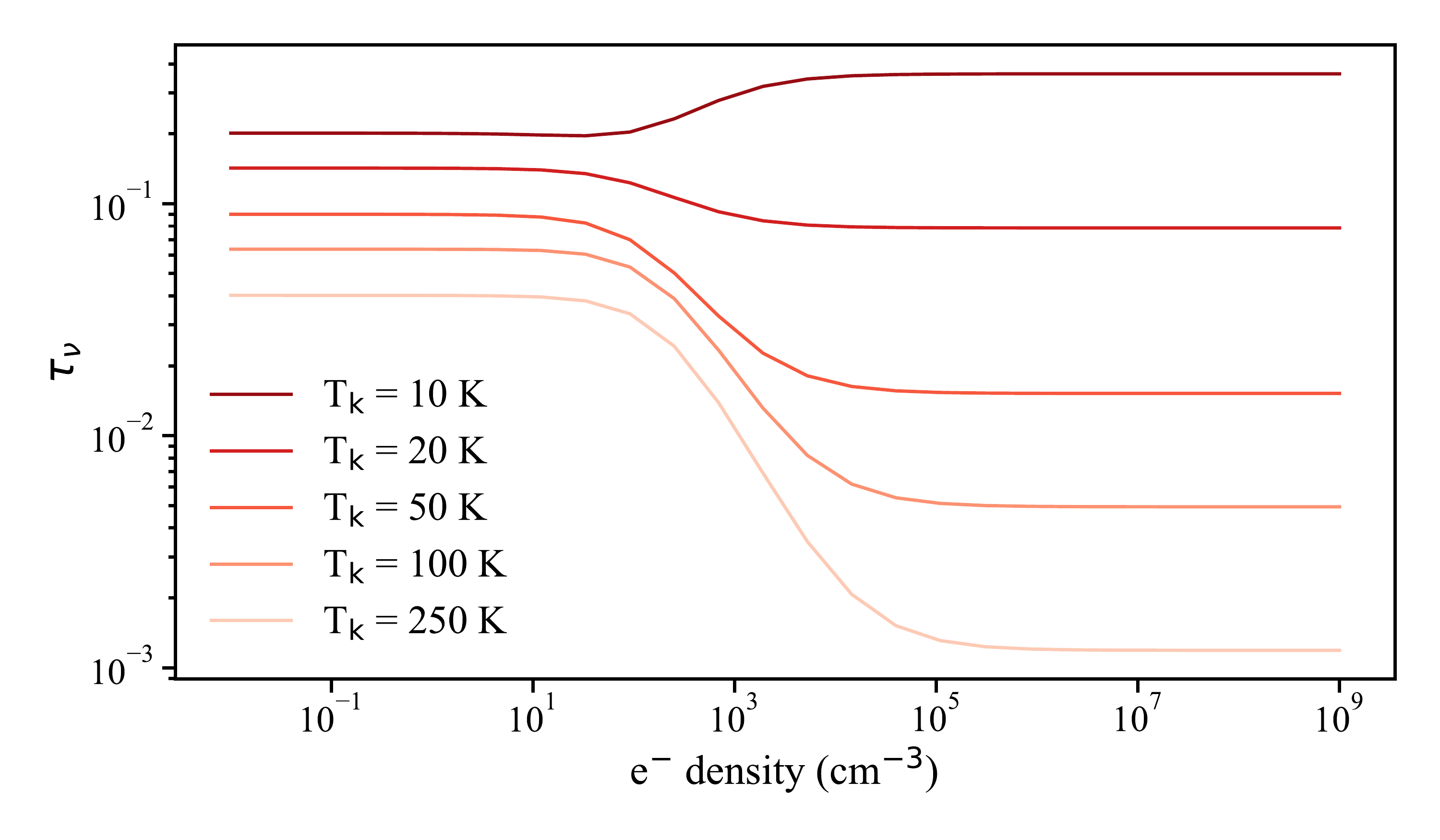}
\caption{(Top) Total CN mass as a function of collider density (with primary collider assumed to be electrons) and kinetic temperature.  The total mass of CN from these models is $(6.2-23) \times 10^{-9}$ M$_{\oplus}$. (Bottom) Optical depth as a function of collider density and kinetic temperature.  In order to check the assumption that CN was optically thin in the debris disk, the calculated CO column densities through the disk midplane were re-scaled by the CN/CO total mass ratio {to obtain a CN column density. This was then turned into an optical depth using Eq.~3 in \cite{2017luca}.}  }\label{fig:luca}
\end{figure}

We begin by converting the upper limit on the CN flux of {25}\,mJy\,km\,s$^{-1}$ to a mass upper limit using Eq. 2 from \citet{2015matranogasfo}.  This calculation assumes that any CN emission at millimeter wavelengths is optically thin, which we will verify {\it a posteriori}.  
To calculate the excitation of the molecule and hence the upper level fractional population, $x_j$, we use the NLTE code developed in the same paper, where $x_j$ will depend on the density of collisional partners (e.g., electrons) and the gas kinetic temperature, $T_k$.  Fig.~\ref{fig:luca} {(top panel) shows the derived total CN masses} from the measured upper limit as a function of collider density and temperature.  The excitation calculation includes fluorescence due to absorption of stellar and interstellar optical/UV photons and later re-emission at lower-energy transitions \citep{2018mat}.  For the stellar radiation field, we assume a PHOENIX stellar model with $T_\mathrm{eff}$=8800\,K, $R_*$=1.7\,R$_\odot$, $\log(g) = 4.0$, $[M/H]=0.06$ (G. Kennedy, private communication), and scaled its flux to what would be received by a CN molecule at 140\,au.  This analysis leads to mass upper limits ranging between {(6.2-23)}$\times 10^{-9}$\,M$_\earth$. 

We then check whether the resulting emission is optically thin along the line of sight to Earth.  {We rescale the CO radial column density from the secondary gas model of \citet{Hughes2017} (measured from the star to the disk's critical radius, and multiplied by two to account for the entire radial extent of the disk) by the CN/CO total abundance ratio to get the CN radial column density,} which we assume is comparable to the column density of CN along the line of sight to Earth.  These assumptions result in CN column densities of {(1.0-3.9)}$\times 10^{12}$\,cm$^{-2}$, depending on excitation, which (via Eq.~3 in \cite{2017luca}, assuming line widths produced by thermal broadening by $T_K$) lead to optical depths of {0.001-0.36} (bottom panel of Fig.~\ref{fig:luca}).  Therefore, CN emission, even if just below our upper limit, would be {largely} optically thin. {Note that this calculation} does not account for uncertainties in the amount of stellar radiation, or the radius at which it is received (140\,au is assumed), or the distance to the star from Earth.  

From there, we can investigate CN photodissociation. By rescaling the previously calculated CO column density along the {vertical direction}, we obtain CN column densities of {(2.2-8.0)$\times 10^{10}$\,cm$^{-2}$},  for which self-shielding is negligible\footnote{https://home.strw.leidenuniv.nl/$\sim$ewine/photo/data/photo\_data/line\_shielding\_functions/CN/photodissociation\_ISRF}. The shielding factors due to CO, H$_2$, and CI column densities \citep[taking CI/CO = 54 from][]{2017higatomic} are {$\sim$0.91, $\sim$0.98 and $\sim$0.43, giving a combined shielding factor of {$\sim$0.38} \citep{2017Heays}}. {Note that} the combined shielding factor is approximated as the product of the shielding factors of the individual species, which assumes that the wavelength regimes over which the shielding takes place do not overlap, but we consider this a reasonable approximation especially given the minor contributions from CO and H$_2$.  Considering only the {ISRF UV field impinging on the disk from the vertical direction}, we obtain a photodissociation timescale for CN of order {$\sim$161} years. 

\subsubsection{Constraints on the HCN/(CO+CO$_2$) outgassing rate ratios}
\label{HCN_rate}

{If we assume that 1) all CN gas is produced by photodissociation of exocometary HCN gas, which is itself produced by cometary release of HCN ice; 2) all CO gas is produced by either cometary release of CO ice, or photodissociation of CO$_2$ released from CO$_2$ ice; and 3) that the production and destruction rates of each species are balanced (i.e., we are in steady state, although this is not necessarily the case in young systems, see \citet{marino2020}), we can relate the observed CN and CO gas masses to the HCN/(CO+CO$_2$) outgassing rate ratio. From Eq. 3 in \citet{2018mat}, this can be expressed as }
\begin{equation}
\label{HCNoverCOoutgassing}
\frac{M_{\rm CN}}{M_{\rm CO}}=\frac{\tau_{\rm CN, phd}^-m_{\rm CN}}{\tau_{\rm CO, phd}^-m_{\rm CO}}\frac{\dot{N}_{\rm HCN}^+}{\dot{N}_{\rm CO+CO_2}^+},
\end{equation}
{where $m_{\rm i}$ is the molecular mass in kg, $\dot{N_{\rm i}^+}$ its production (outgassing) rate in number of molecules per second,  $M_{\rm i}$ the observed gas mass, and $\tau_{\rm i}^-$ its photodissociation timescale of the gas species.}

{The observed gas masses $M_{\rm CO}$ and $M_{\rm CN}$ are well constrained observationally from \citet{moor2019} and our data, and the CN photodissociation timescale $\tau_{\rm CN, phd}^-$ is also relatively well constrained, as shielding only increases it by a factor of a few (Sect. \ref{CNphd}). On the other hand, the main uncertainty in the calculation remains the CO photodissociation timescale $\tau_{\rm CO, phd}^-$, which can range from a few hundred years { if shielding by CI is neglected} to much longer than the system age of 40 Myr.}

{Taking the CO photodissociation timescale $\tau_{\rm CO, phd}^-\sim${3,300} yr derived in our second-generation, CI-free model} would imply a HCN/(CO+CO$_2$) outgassing rate ratio of {(1.2-4.6)} $\times 10^{-5}$. This is at least a factor of $\sim${65} below the observed values for Solar system comets, ranging from about $3\times10^{-3}-2$ \citep[see Fig.~5 in][]{2018mat}. Therefore, in the absence of CI shielding the constraint on the flux density of CN and HCN implies that either CO may be preferentially released from the exocometary ice relative to HCN, or there may be a true HCN depletion compared to CO ice in 49 Ceti's exocomets.

Perhaps more likely, however, given the work of \citet{2018kral} and the observed presence of CI, is that CI is preferentially shielding CO from UV photodissociation. For example, if we assume the CI/CO ratio of 54 derived from \citet{2017higatomic} (which does not account for optical depth and excitation effects) we obtain a CO shielding factor {of $\sim10^{-15}$. This is} so small that it would imply that CO is hardly photodissociating at all, allowing it to survive for much longer than the age of the system.  Shielding by CI has a much weaker effect on other molecules, since the CO photodissociation bands are all at wavelengths shorter than the CI ionization threshold (110 nm), whereas some photodissociation bands of CN (and those of most other cometary molecules) extend beyond 110\,nm, where the stellar radiation field is orders of magnitude stronger.  

It therefore seems likely that the only reason CO appears so anomalously abundant is due to shielding by CI, and that CN is in fact photodissociated about as rapidly as we would expect given the harsh UV radiation field from both the ISRF and the central A star. {Whether CI shielding is the cause of the overabundance of CO compared to other molecules could be confirmed with an accurate measurement of the CI column density. This can be achieved with future high resolution, optically thin, multi-transition CI observations, which are best suited to solving the degeneracy between excitation, spatial structure, mass, and optical depth which affects single line measurements. If CI column densities are confirmed to be very high,} then the CO photodissociation timescale $\tau_{\rm CO, phd}^-$ in Eq. \ref{HCNoverCOoutgassing} could be orders of magnitude higher than derived in the absence of CI shielding, and allow us to reconcile our extreme CN/CO flux ratios around 49 Ceti with Solar System comet -like HCN/(CO+CO$_2$) ice abundances.

%
\section{Conclusions}
\label{section: conclusion}

In order to constrain the composition of the gas in nearby debris disks, we performed a deep survey for five molecular species in the disk around 49 Ceti.  After 3.2 hours of integration with ALMA at low angular and spectral resolution to maximize sensitivity, and after applying a spatial-spectral shifting technique adapted from \citet{2017fomalhaut}, we obtain stringent upper limits { from the shifted and velocity-range-restricted moment 0 maps} that represent the most sensitive search to date for molecules other than CO in a debris disk: { HCN(4-3) at 31\,mJy\,km\,s$^{-1}$, HCO$^{+}$(4-3) at 37\,mJy\,km\,s$^{-1}$, CN(3-2) at 25\,mJy\,km\,s$^{-1}$, SiO(8-7) at 34\,mJy\,km\,s$^{-1}$, CH$_3$OH 1(1,1)-0(0,0) at 22\,mJy\,km\,s$^{-1}$, and CH$_3$OH 4(1,3)-3(0,3) at 56\,mJy\,km\,s$^{-1}$}.   

We use our upper limits on the flux density to calculate flux ratios that we can compare with the composition of protoplanetary disks and solar system comets.  The upper limits on the HCN/CO and HCO$^+$/CO line ratios are inconsistent with a protoplanetary composition for the disk material.  The upper limits on the HCN/CO ratio and the nondetection of HCO$^+$ and methanol are also inconsistent with the range of cometary compositions observed within the Solar System.  Since the temperatures and mechanisms involved in the release of exocometary gas at tens to hundreds of au from the central star may differ markedly from those involved in the au-scale sublimation of Solar System comets, we turn to models of second-generation gas production to resolve this apparent conundrum.  While the CO photodissociation timescale is apparently much shorter than the age of the system, it can be extended by orders of magnitude by taking into account shielding by CI \citep{2018kral}, which is observed to be present in abundance in the 49 Ceti system, likely as a byproduct of CO photodissociation \citep{2017higatomic}.  Using a NLTE code from \citet{2018mat}, we demonstrate that if CI shielding is neglected, then the implied HCN/CO outgassing ratio for 49 Ceti would be orders of magnitude lower than the range observed for Solar System comets.  

Our results imply that either there is a marked difference in the chemical abundance of the 49 Ceti system relative to all known Solar System comets and protoplanetary disks, an unspecified mechanism causing preferential release of CO relative to HCN, or more plausibly, that our results provide support for a picture where CI shields CO, but cannot shield CN and other (exo-)cometary molecules.  The apparently high abundance of CO relative to the other molecules therefore may not reflect the true chemical abundances of the icy bodies in the system, but rather reflects the fact that only CO is shielded from photodissociation and therefore accumulates in substantial amounts while the other molecules are photodissociated so rapidly as to be effectively invisible.  Our deep limits on molecular species readily detected in protoplanetary disks and comets provide support for the  shielding of CO by CI in the 49 Ceti system.

%
\section*{Acknowledgments}

We thank Stew Novick for help with deciphering the CDMS database, and Grant Kennedy for providing the stellar photospheric model fits.  Support for this work was provided by the NSF through award SOSPA5-009 from the NRAO.  A.M.H. is supported by a Cottrell Scholar Award from the Research Corporation for Science Advancement.  A.M. acknowledges the support of the Hungarian National Research, Development  and  Innovation  Office  NKFIH  Grant  KH-130526. 
LM acknowledges support from the Smithsonian Institution as a Submillimeter Array (SMA) Fellow.  The work of JK is supported by NOIRLab, which is managed by the Association of Universities for Research in Astronomy (AURA) under a cooperative agreement with the National Science Foundation.
This paper makes use of the following ALMA data:  ADS/JAO.ALMA\#2017.1.00941.S 
ALMA is a partnership of ESO (representing its member states), NSF (USA) and NINS (Japan), together with NRC (Canada), MOST and ASIAA (Taiwan), and KASI (Republic of Korea), in cooperation with the Republic of Chile.  The Joint ALMA Observatory is operated by ESO, AUI/NRAO and NAOJ.  
The National Radio Astronomy Observatory is a facility of the National Science Foundation operated under cooperative agreement by Associated Universities, Inc.
This work has made use of data from the European Space Agency (ESA) mission {\it Gaia} (\url{https://www.cosmos.esa.int/gaia}), processed by the {\it Gaia}
Data Processing and Analysis Consortium (DPAC, \url{https://www.cosmos.esa.int/web/gaia/dpac/consortium}). 
Funding for the DPAC has been provided by national institutions, in particular the institutions participating in the {\it Gaia} Multilateral Agreement.  

\software{
\texttt{CASA}  \citep{2007mcmull},
\texttt{MIRIAD} \citep{sault95},
\texttt{NumPy} \citep{van2011numpy}, 
\texttt{Astropy} \citep{astropy},  
\texttt{Pandas} \citep{mckinney}, 
\texttt{Uncertainties}, \url{http://pythonhosted.org/uncertainties/} 
}

\clearpage
\bibliography{bibliography.bib}
\end{document}